\documentclass[10pt,journal,twocolumn,letterpaper]{IEEEtran}

\usepackage{amsmath,amsfonts,amssymb}
\usepackage{algorithmic}
\usepackage{algorithm}
\usepackage{array}
\usepackage{tabularx}
\usepackage{multirow}

\usepackage{url}
\usepackage{graphicx}
\usepackage{cite}
\usepackage{bm}
\usepackage{booktabs}
\usepackage[T1]{fontenc}
\usepackage{newtxmath}
\begin{document}
\pagestyle{plain}

\title{StarWorld: A Cross-Layer World Model for Starlink Prediction and Streaming}
\author{Bowen Zhang, Pengcheng Luo, Hongjun Xie, Shujin Jia, Genke Yang, and Jian Chu\thanks{This work was supported by the National Science and Technology Major Project of China under Grant 2025ZD1602400. \emph{(Corresponding author: Genke Yang.)}}
\thanks{Bowen Zhang, Pengcheng Luo, Hongjun Xie, Genke Yang, and Jian Chu are with Ningbo Artificial Intelligence Institute, Shanghai Jiao Tong University, Ningbo 315000, China, and also with the School of Automation and Intelligent Sensing, Shanghai Jiao Tong University, Shanghai 200240, China, and the Key Laboratory of System Control and Information Processing, Ministry of Education of China, Shanghai 200240, China (e-mail: bwz96sco@sjtu.edu.cn, luopeng69131@sjtu.edu.cn, xiehongjun@sjtu.edu.cn, gkyang@sjtu.edu.cn, chujian@niii.com).}
\thanks{Hongjun Xie is also with Shanghai i-Space Orbital Computing Infrastructure Technology Co., Ltd., Shanghai 200235, China.}
\thanks{Shujin Jia is with the Research Institute, Baoshan Iron \& Steel Co., Ltd., Shanghai 201999, China (e-mail: jiashujin@baosteel.com).}
\thanks{The source code will be available at: \protect\url{https://github.com/luopeng69131/StarWorld}.}
}

\maketitle
\thispagestyle{plain}

\begin{abstract}
Starlink and other low Earth orbit (LEO) satellite broadband systems exhibit stochastic access dynamics whose effects propagate from network measurements to throughput prediction and adaptive bitrate (ABR) streaming, making reasoning over multiple plausible futures essential. Existing studies mainly characterize realized traces or individual system layers, leaving cross-layer world modeling of these evolving dynamics largely unexplored. To address this gap, we propose \emph{StarWorld}, a cross-layer world model for Starlink prediction and streaming. StarWorld couples learned stochastic network dynamics with executable prediction, control, and playback processes to imagine multi-step cross-layer futures, while a calibrated Trust Gate determines whether these futures are sufficiently reliable for use. On real Starlink measurements, StarWorld achieves the lowest network-future distributional errors among the comparison methods at the principal horizons and reduces pairwise controller quality-of-experience (QoE) effect error by 76\%. These results support StarWorld as a cross-layer imagination environment for studying prediction and streaming outcomes under uncertain Starlink access conditions.
\end{abstract}


\section{Introduction}
\label{sec:introduction}

\subsection{Background and Motivation}

\IEEEPARstart{S}{tarlink} and other low Earth orbit (LEO) satellite broadband systems are emerging as a new access infrastructure for remote, maritime, and mobile environments~\cite{mohan2024multifaceted}. End-user measurements reveal substantial variation in throughput and latency across access conditions~\cite{ma2023network}. Multimedia studies further show that satellite switching and buffer configuration affect playback continuity and service quality~\cite{zhao2023realtime}. These findings motivate a cross-layer view of Starlink performance that links access dynamics with throughput prediction and streaming behavior.

\begin{figure*}[t]
  \centering
  \includegraphics[width=\textwidth]{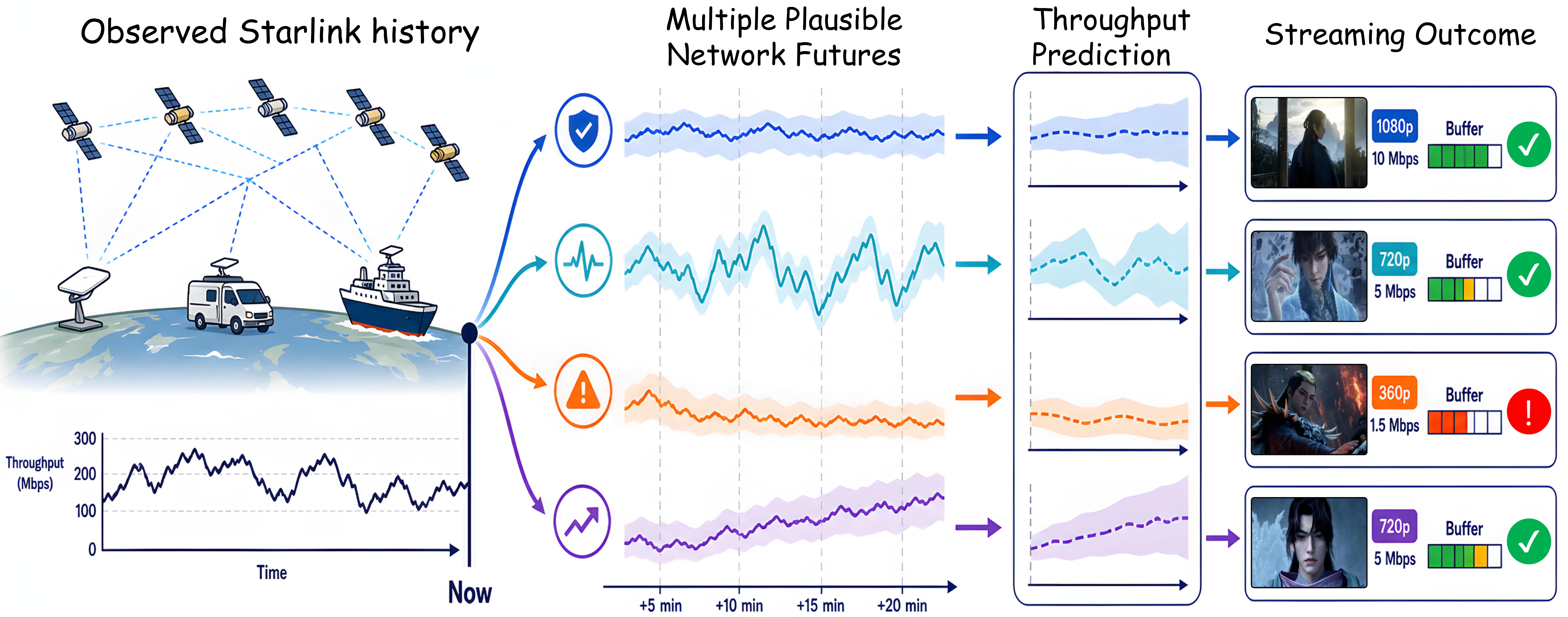}
  \caption{Starlink cross-layer uncertainty: one observed history can lead to different network, prediction, and streaming outcomes.}
  \label{fig:starworld-motivation}
\end{figure*}

Such cross-layer evaluation requires accounting for temporal dependence, uncertainty, and their effects on application behavior. Starlink throughput exhibits structure at multiple timescales~\cite{garcia2023multi}, and prediction models use measurement histories to estimate future capacity~\cite{liu2025starnet}. Streaming controllers then combine capacity estimates with buffer state to select bitrates, coupling network conditions to playback quality~\cite{yin2015control,yuan2024jointqoe}. Within the user-side scope, future access capacity is treated as an exogenous process that is independent of application actions. Application actions instead determine prediction and playback outcomes under a given capacity evolution. A single replay evaluates these interactions on one realized network trajectory; it does not characterize the range of plausible continuations or their associated playback risks. Understanding these effects therefore requires keeping uncertain network evolution and its prediction and streaming consequences connected across the same multi-step futures, as illustrated in Fig.~\ref{fig:starworld-motivation}.

Existing studies provide complementary foundations for cross-layer evaluation. Starlink measurement campaigns characterize access variation across locations, timescales, and mobility states~\cite{michel2022first,kassem2022browser,ma2023network,mohan2024multifaceted}. Prediction models forecast link capacity from measured histories~\cite{liu2025starnet,lv2024accurate}, while streaming experiments assess playback outcomes along supplied network trajectories~\cite{zhao2023realtime,meng2021pitree}. Data-driven network models predict delay, jitter, and loss under specified traffic and routing configurations~\cite{ferriol2023routenetfermi}, and hardware-assisted learning supports performance estimation using physical testbed measurements~\cite{guemes2026routenetgauss}. These studies establish access characterizations, capacity forecasts, and performance estimators, but generally target individual quantities or evaluate applications on supplied network trajectories. What remains less developed is a joint dynamical representation that generates coherent network, prediction, and playback trajectories across multiple plausible continuations of the same observed history.

World models provide a natural starting point: they represent environment states and learn transition dynamics that can be rolled forward to generate multi-step futures~\cite{chua2018pets,hafner2019planet,hafner2020dreamer,schrittwieser2020muzero}. A Starlink cross-layer system contains two distinct forms of dynamics: user-observable network evolution is stochastic and difficult to specify analytically, whereas predictors, controllers, and playback processes expose executable interfaces or well-defined transition semantics. Physics-informed learning combines learned models with known system structure~\cite{karniadakis2021physicsinformed}. Network digital twins further connect virtual representations with physical observations to support wireless decisions~\cite{salehi2024multiverse,li2025networktwins}. Together, these ideas suggest a hybrid cross-layer world model that learns uncertain network evolution, preserves executable application semantics, and creates an imagination environment spanning measurement, prediction, and streaming. Because the reliability of imagined futures varies with uncertainty, data support, and rollout length, this environment also requires an explicit trust boundary.

\begin{figure*}[t]
  \centering
  \includegraphics[width=\textwidth]{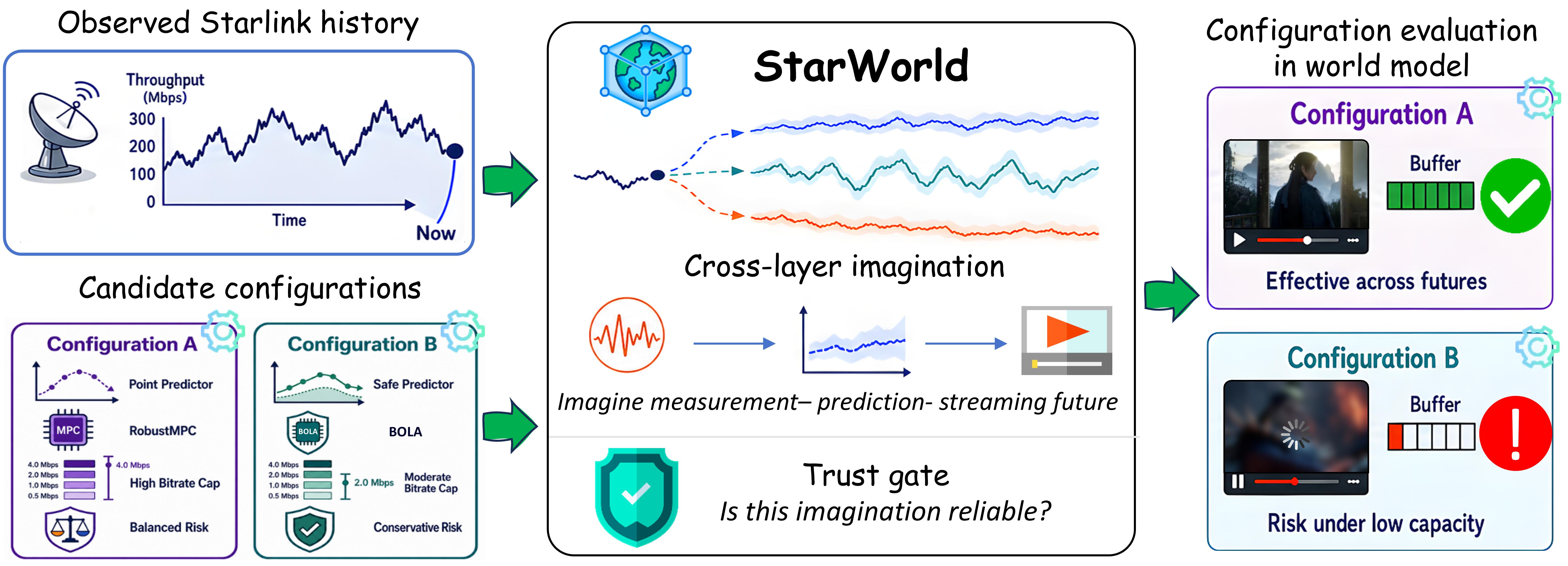}
  \caption{StarWorld concept: imagining shared cross-layer futures and comparing configuration outcomes within a calibrated trust boundary.}
  \label{fig:starworld-concept}
\end{figure*}

Motivated by this perspective, we propose \emph{StarWorld}, a cross-layer world model for Starlink prediction and streaming. StarWorld learns stochastic network evolution and connects the resulting futures with executable prediction, control, and playback dynamics to generate coherent multi-step trajectories across the three layers. A Trust Gate identifies the reliability boundary of these imagined futures. StarWorld thereby provides an imagination environment for studying how uncertain Starlink access conditions evolve and propagate across system layers. Fig.~\ref{fig:starworld-concept} illustrates this high-level concept.

\subsection{Contributions}

The main contributions of this paper are as follows.
\begin{itemize}
    \item We propose StarWorld, a cross-layer world model for Starlink prediction and streaming. It provides an imagination environment that rolls uncertain Starlink access conditions forward into coherent network, prediction, and playback futures.

    \item We develop a hybrid learned--executable design with Cross-Layer Imagination and a Trust Gate. The design preserves application semantics across probabilistic rollouts and identifies the domain and horizon over which imagined futures can be used reliably.

    \item We evaluate StarWorld on real Starlink measurements and cross-layer replay outcomes. It obtains the lowest throughput and lower-tail distributional errors at the principal horizons and reduces controller quality-of-experience (QoE) effect error by 76\%, while providing near-perfect unsupported-query detection. 
\end{itemize}

\subsection{Organization}

The remainder of this paper is organized as follows. Section~II reviews Starlink cross-layer system evaluation and research on world models and trustworthy imagination. Section~III formulates the Starlink cross-layer world-modeling problem, and Section~IV presents Cross-Layer Imagination and the Trust Gate. Section~V describes the experimental setting and evaluation results. Section~VI concludes the paper and discusses future work.

\section{Related Work}
\label{sec:related-work}

This section reviews two complementary bodies of research. The first establishes the empirical and executable foundations for modeling Starlink dynamics across measurement, prediction, and streaming. The second examines probabilistic world modeling, hybrid learned--executable structures, and the reliability of imagined trajectories.

\subsection{Starlink Dynamics and Cross-Layer System Evaluation}

A useful Starlink system model must first identify which user-side access states vary over time. Early measurement studies demonstrated practically useful throughput and latency performance~\cite{michel2022first,kassem2022browser}. Subsequent end-user, multi-timescale, and cross-regional campaigns exposed throughput bursts, latency variation, transient disruptions, and context-dependent behavior~\cite{ma2023network,garcia2023multi,mohan2024multifaceted}. These findings show that Starlink access cannot be characterized adequately by a single average rate; it exhibits dynamic structure conditioned on recent history and operating context. They provide an empirical basis for stateful modeling, while also distinguishing retrospective trace statistics from forecasting future evolution from a current history.

Research on Starlink prediction has begun to model these dynamics at different temporal scales. StarNet learns throughput sequences from long-term measurements, fine-grained models characterize and predict burst-level transmission behavior, and Horizon forecasts throughput and latency at regional and hourly scales~\cite{liu2025starnet,garcia2025modeling,benghe2026horizon}. In adaptive video streaming, Lumos and Fugu integrate throughput or transmission-time forecasts into bitrate decisions, demonstrating that a capacity estimate acquires operational meaning through application behavior~\cite{lv2024accurate,yan2020fugu}. Together, these studies provide sequence models, probabilistic forecasts, and application-facing capacity signals. Extending these capabilities into a cross-layer world model requires uncertain network futures to remain connected to their prediction and playback consequences.

Streaming research provides the executable mapping from a network trajectory to application outcomes. Model predictive control (MPC)-based adaptive bitrate methods use throughput estimates and playback-buffer state for finite-horizon decisions~\cite{yin2015control}. Starlink multimedia measurements show that handovers and capacity variation affect video on demand, live streaming, and interactive video in different ways~\cite{zhao2023realtime}, while studies of low-latency streaming using Dynamic Adaptive Streaming over HTTP (DASH) compare control strategies in real systems or simulation~\cite{zhao2024dash}. Large-scale video measurements further reveal how Starlink variability propagates into bitrate adaptation and rebuffering~\cite{izhikevich2024global}. Together, these studies provide measured trajectories, forecasting models, and executable streaming components. Their integration around multiple plausible network futures would connect uncertain access evolution with its prediction and playback consequences in a common imagination environment.

\subsection{World Models and Trustworthy Cross-Layer Imagination}

World-model research provides a general foundation for such an environment. Probabilistic ensembles with trajectory sampling (PETS) generate uncertain action-conditioned futures from learned probabilistic dynamics~\cite{chua2018pets}. PlaNet and Dreamer combine deterministic memory with stochastic latent states in compact representations and recursively imagine multi-step action outcomes~\cite{hafner2019planet,hafner2020dreamer}. MuZero shows that a learned model can preserve the task semantics needed for planning without reconstructing every detail of the environment, and DreamerV3 extends world-model learning across diverse control tasks~\cite{schrittwieser2020muzero,hafner2025dreamerv3}. This progression shifts the role of a world model from next-state prediction toward an internal environment for imagining decision-relevant long-horizon futures.

Multi-step imagination also introduces model bias and data-support concerns. Errors can accumulate with rollout length, and alternative actions may move the imagined state beyond the training distribution. Model-based policy optimization (MBPO) balances synthetic-data gains against model bias by launching short model rollouts from real states, whereas model-based offline policy optimization (MOPO) uses model uncertainty to treat offline distribution shift conservatively~\cite{janner2019mbpo,yu2020mopo}. Deep ensembles provide a scalable empirical estimate of predictive uncertainty, and selective prediction casts abstention as a risk--coverage tradeoff~\cite{lakshminarayanan2017deepensembles,geifman2019selectivenet}. Generating a trajectory therefore does not imply equal reliability across horizons or query conditions. Trustworthy imagination requires independently calibrated uncertainty, explicit joint-support checks, and selective rejection beyond the validated rollout horizon.

A further foundation for domain-specific world models is the combination of known system semantics with learned dynamics. Physics-informed machine learning couples data-driven models with established equations and structural constraints~\cite{karniadakis2021physicsinformed}. Network digital twins likewise combine data-driven representations with operational system structure for simulation and system analysis~\cite{li2025networktwins}. Taken together, these principles leave open how to construct a Starlink cross-layer world model in which uncertain user-side access evolution is learned, prediction and streaming semantics remain executable, and imagined futures are bounded by calibrated uncertainty, joint query support, and rollout horizon.

\section{Cross-Layer World Modeling Problem}
\label{sec:problem-formulation}

A recorded Starlink trajectory reveals the outcome of a predictor or streaming configuration only on one realized network path. It does not show which futures may follow the current state, how candidate configurations behave across those futures, or whether the resulting estimates remain within a reliable modeling range. We formulate this requirement as a cross-layer what-if evaluation problem and define its query, output, and causal scope below.

\subsection{What-If Evaluation under Uncertain Starlink Futures}

Future Starlink access conditions are stochastic: similar histories can develop into stable, volatile, or lower-capacity continuations. The same streaming configuration can consequently produce different prediction errors, bitrate decisions, buffer trajectories, and QoE outcomes. Evaluating a candidate configuration therefore requires a distribution over plausible futures.

Accordingly, the evaluation is organized around three questions:
\begin{enumerate}
  \item \emph{Which network futures remain plausible from the current cross-layer history?}
  \item \emph{How does each candidate configuration behave across those shared futures?}
  \item \emph{Is the resulting configuration comparison sufficiently supported for use?}
\end{enumerate}
Answering these questions requires configuration outcome distributions and paired effects, together with an explicit decision to release the comparison or abstain.

\subsection{Query, Output, and Modeling Scope}

\begin{figure*}[t]
  \centering
  \includegraphics[width=\textwidth]{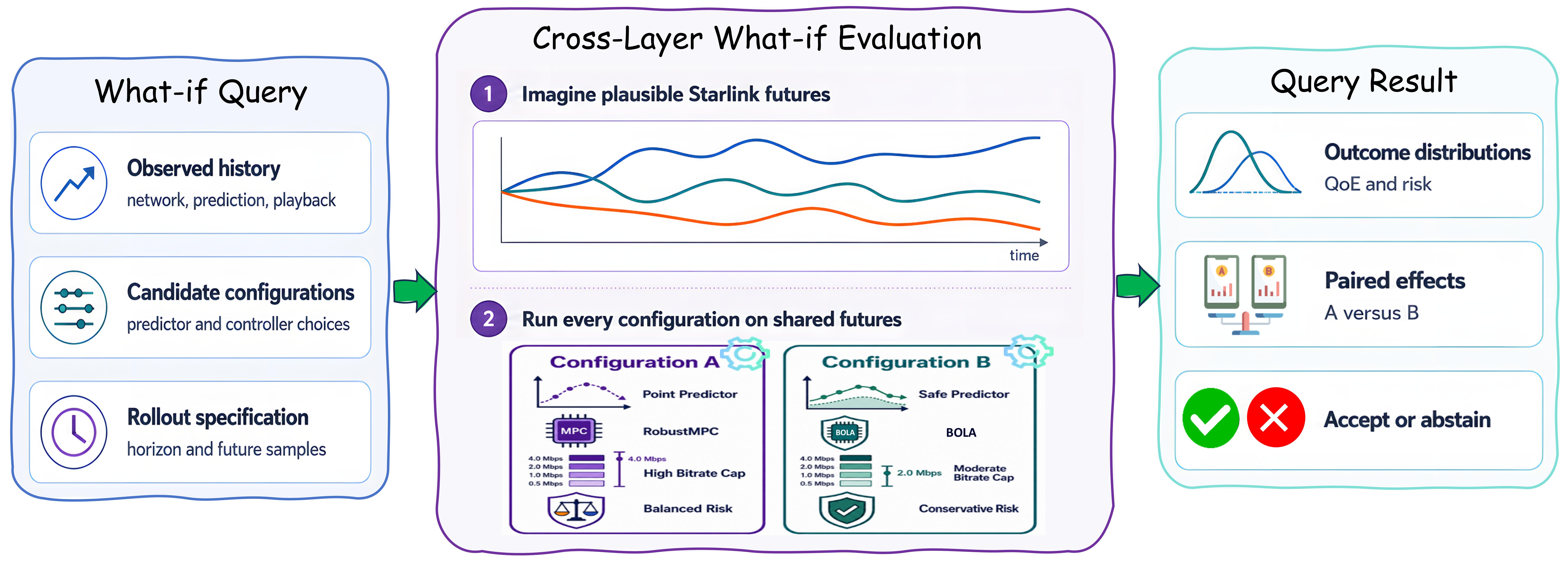}
  \caption{StarWorld query interface: observed history and candidate configurations mapped to outcome distributions, paired effects, and a trust decision.}
  \label{fig:starworld-problem-interface}
\end{figure*}

A StarWorld what-if evaluation is specified by its available observations, candidate configurations, returned result, and causal scope. These elements determine what information may enter a query, what comparison may be released, and which system effects the model is intended to represent. Fig.~\ref{fig:starworld-problem-interface} summarizes this query--evaluation--result interface.

A StarWorld query contains only information available at the current time. Let
\begin{equation}
\bm{s}_{\leq t}
=
\left(
\bm{x}_{\leq t},
\bm{p}_{\leq t},
\bm{b}_{\leq t}
\right)
\label{eq:observed-cross-layer-history}
\end{equation}
denote the observed cross-layer history, where $\bm{x}$, $\bm{p}$, and $\bm{b}$ represent network, prediction, and playback states, respectively. Using the history in \eqref{eq:observed-cross-layer-history}, a query is defined as
\begin{equation}
\mathcal{Q}_{t}
=
\left(
\bm{s}_{\leq t},
\bm{c},
\mathcal{U},
H,
M
\right),
\label{eq:starworld-query}
\end{equation}
where $\bm{c}$ denotes observed exogenous context, $\mathcal{U}=\{\bm{u}_{j}\}_{j=1}^{J}$ is the set of candidate predictor--controller configurations, $H$ is the rollout horizon, and $M$ is the number of sampled futures. Future throughput, future prediction error, and complete replay outcomes are excluded from the query.

For each configuration $\bm{u}_{j}$, StarWorld generates $M$ measurement--prediction--streaming trajectories $\{\widehat{\tau}_{j}^{(m)}\}_{m=1}^{M}$. These trajectories induce an outcome distribution $\widehat{P}(\mathcal{Y}_{j})$ for a QoE or risk functional $\mathcal{Y}$ and paired configuration effects $\Delta_{j,\ell}^{(m)}$. Cross-Layer Imagination collects them into the internal result package
\begin{equation}
\mathcal{Z}_t
=
\left(
\widehat{\mathcal{T}},
\widehat{\mathcal{P}}_{\mathcal{Y}},
\widehat{\bm{\Delta}}
\right),
\label{eq:imagination-result-package}
\end{equation}
where $\widehat{\mathcal{T}}$ contains the sampled cross-layer trajectories, $\widehat{\mathcal{P}}_{\mathcal{Y}}$ contains their outcome distributions, and $\widehat{\bm{\Delta}}$ contains the paired configuration effects. The Trust Gate determines whether this package can be released. Accordingly, the external output is
\begin{equation}
\operatorname{StarWorld}(\mathcal{Q}_{t})
=
\begin{cases}
(\mathcal{Z}_t,\mathrm{accept}), & d_t=\mathrm{accept},\\
(r_t,\mathrm{abstain}), & d_t=\mathrm{abstain},
\end{cases}
\label{eq:starworld-output}
\end{equation}
where $d_t\in\{\mathrm{accept},\mathrm{abstain}\}$ is the Trust Gate decision and $r_t$ identifies the failed trust condition.

The model is restricted to user-observable Starlink access dynamics. Future network capacity is treated as exogenous to application-side configurations and runtime bitrate actions:
\begin{equation}
p\!\left(
\bm{x}_{t+1}\mid
\bm{s}_{\leq t},\bm{c},\bm{u}_{j},\bm{a}_{t}
\right)
=
p\!\left(
\bm{x}_{t+1}\mid
\bm{x}_{\leq t},\bm{c}
\right).
\label{eq:exogenous-network-scope}
\end{equation}
Configurations affect the prediction and playback consequences observed under a network future, but do not alter that future. This scope supports controlled application-side comparisons without making claims about Starlink's internal scheduling or physical mechanisms.

\section{StarWorld Framework}
\label{sec:starworld-framework}

StarWorld answers a query through two core components. Cross-Layer Imagination generates plausible network futures and propagates candidate configurations through prediction and streaming behavior. The Trust Gate then determines whether the imagined comparison lies within the model's validated range. Model learning and gate calibration are described after this query-time workflow.

\subsection{Framework Overview}

\begin{figure*}[t]
  \centering
  \includegraphics[width=\textwidth]{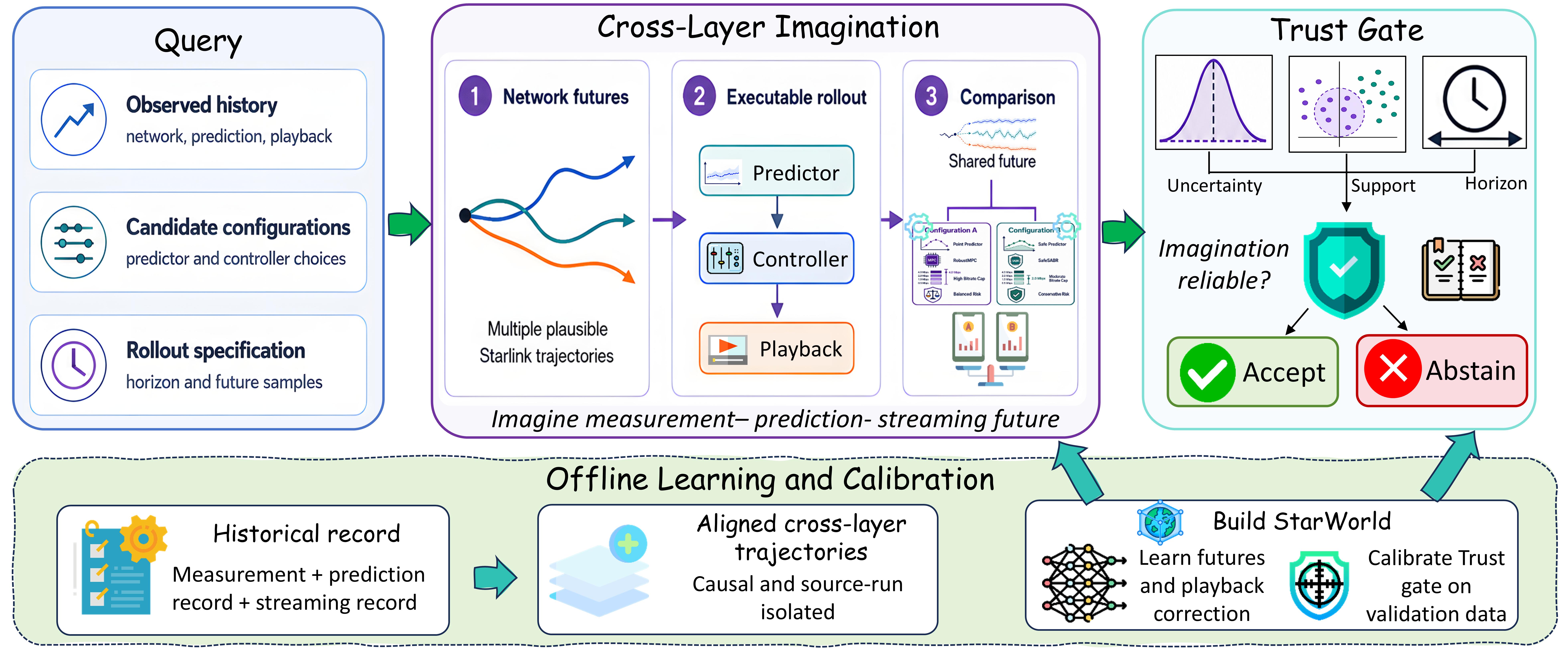}
  \caption{StarWorld framework: Cross-Layer Imagination and the Trust Gate, supported by offline learning and calibration.}
  \label{fig:starworld-framework}
\end{figure*}

Starting from an observed history, StarWorld samples several plausible network continuations. Every candidate configuration is executed on the same sampled futures, producing distributions of prediction, playback, QoE, and risk outcomes. The resulting comparison is released only when its uncertainty, data support, and horizon satisfy the Trust Gate. Fig.~\ref{fig:starworld-framework} presents this workflow and separates query-time operation from offline learning and calibration.

\subsection{Cross-Layer Imagination}

Cross-Layer Imagination determines what network futures may occur and what application outcomes each candidate configuration produces in those futures. It proceeds in three stages: probabilistic network generation, executable cross-layer rollout, and paired comparison under shared futures.

\subsubsection{Probabilistic Network Futures}

StarWorld learns stochastic user-side network dynamics from causal measurement histories. From the same history and exogenous context, it samples
\begin{equation}
\widehat{\bm{x}}^{(m)}_{t+1:t+H}
\sim
\mathcal{F}_{\theta}
\left(
\bm{x}_{\leq t},
\bm{c}
\right),
\qquad m=1,\ldots,M.
\label{eq:probabilistic-network-futures}
\end{equation}
The stochastic ensemble represents both ordinary continuations and lower-tail capacity behavior, preventing multiple plausible futures from being collapsed into a single mean forecast. Each sampled path is temporally coherent and remains independent of the application configuration by the scope in \eqref{eq:exogenous-network-scope}.

\subsubsection{Executable Cross-Layer Rollout}

For every sampled network future, StarWorld invokes the predictor and controller specified by configuration $\bm{u}_{j}$ and advances the playback state. At imagined step $h$, the executable operations are summarized as
\begin{equation}
\begin{aligned}
\widehat{\bm{p}}^{(m,j)}_{t+h}
&=
G_{\bm{u}^{P}_{j}}
\left(
\widehat{\bm{x}}^{(m)}_{\leq t+h}
\right),\\
\widehat{\bm{a}}^{(m,j)}_{t+h}
&=
\pi_{\bm{u}^{S}_{j}}
\left(
\widehat{\bm{b}}^{(m,j)}_{t+h-1},
\widehat{\bm{p}}^{(m,j)}_{t+h}
\right),\\
\widehat{\bm{b}}^{(m,j)}_{t+h}
&=
\mathcal{P}
\left(
\widehat{\bm{b}}^{(m,j)}_{t+h-1},
\widehat{\bm{x}}^{(m)}_{t+h},
\widehat{\bm{a}}^{(m,j)}_{t+h}
\right)
\\
&\quad+
\delta_{\phi}
\left(
\widehat{\bm{s}}^{(m,j)}_{t+h-1},
\widehat{\bm{a}}^{(m,j)}_{t+h}
\right).
\end{aligned}
\label{eq:executable-cross-layer-rollout}
\end{equation}
In \eqref{eq:executable-cross-layer-rollout}, $G$ and $\pi$ retain the executable semantics of the configured predictor and controller, $\mathcal{P}$ is the analytical playback transition, and $\delta_{\phi}$ is a learned correction for replay behavior not captured by the analytical update. The resulting trajectory can be written compactly as
\begin{equation}
\widehat{\tau}^{(m)}_{j}
=
\operatorname{Imagine}
\left(
\bm{s}_{\leq t},
\widehat{\bm{x}}^{(m)}_{t+1:t+H},
\bm{u}_{j}
\right).
\label{eq:cross-layer-imagination}
\end{equation}
The configuration remains fixed within a rollout, whereas the controller produces runtime actions from the evolving imagined state.

\subsubsection{Shared-Future Configuration Comparison}

All candidate configurations reuse the same indexed network futures. Holding network randomness common across branches makes their outcome differences reflect predictor, controller, and runtime-action behavior. For configurations $j$ and $\ell$, their paired effect in future $m$ is
\begin{equation}
\Delta^{(m)}_{j,\ell}
=
\mathcal{Y}
\left(
\widehat{\tau}^{(m)}_{j}
\right)
-
\mathcal{Y}
\left(
\widehat{\tau}^{(m)}_{\ell}
\right).
\label{eq:shared-future-effect}
\end{equation}
Aggregating the paired effects in \eqref{eq:shared-future-effect} yields both the expected difference and its variability across plausible network futures. Cross-Layer Imagination therefore produces future distributions, configuration outcome distributions, and paired effects for subsequent trust assessment.

\subsection{Trust Gate}

The Trust Gate defines the usable boundary of a StarWorld result. It operates after Cross-Layer Imagination and determines whether a generated cross-layer comparison can be released or whether StarWorld should abstain. For each comparison, the gate follows three checks. It first examines disagreement and predictive spread across the imagined futures, then determines whether the observed context, candidate configurations, and generated actions remain covered by the training data, and finally checks whether the requested rollout length lies within the validated horizon. The comparison is released only when all three checks pass; otherwise, the failed check becomes the abstention reason.

Formally, these checks use both the original query and the Cross-Layer Imagination result. The Trust Gate input is
\begin{equation}
\mathcal{I}^{\mathrm{gate}}_t
=
\left(\mathcal{Q}_t,\mathcal{Z}_t\right).
\label{eq:trust-gate-input}
\end{equation}
Here, the query contributes the observed state, context, candidate configurations, and requested horizon, while the result package in \eqref{eq:imagination-result-package} contributes the sampled futures, generated action sequences, outcome distributions, and paired effects. From this combined input, the gate derives rollout uncertainty $U_t$, joint data-support distance $D_t$, and the horizon condition formalized below.

Rollout uncertainty is obtained from the probabilistic futures in $\widehat{\mathcal{T}}$. Let $\bm{\mu}_{e,h}$ and $\bm{\sigma}_{e,h}^{2}$ denote the network mean and element-wise aleatoric variance predicted by ensemble member $e$ at imagined step $h$, and let $d_x$ be the network-state dimension. StarWorld combines disagreement among members with their predicted variance as
\begin{equation}
U_t^2
=
\frac{1}{H d_x}
\sum_{h=1}^{H}
\left(
\operatorname{tr}\!\left(
\operatorname{Cov}_{e}[\bm{\mu}_{e,h}]
\right)
+
\mathbb{E}_{e}\!\left[
\lVert\bm{\sigma}_{e,h}\rVert_2^2
\right]
\right),
\label{eq:rollout-uncertainty}
\end{equation}
and $U_t$ is its nonnegative square root. This score increases when model members disagree or when an individual member predicts a diffuse future.

Data support is evaluated on the complete cross-layer query exercised by each candidate. For configuration $j$, the support vector is
\begin{equation}
\bm{z}_{t,j}
=
\left[
\rho(\bm{s}_{\leq t},\bm{c}),
\bm{u}_{j},
\psi(\widehat{\bm{a}}_{j,1:H})
\right],
\label{eq:joint-support-vector}
\end{equation}
where $\rho(\cdot)$ summarizes the observed state and context and $\psi(\cdot)$ summarizes the action sequence generated during the executable rollout. With training-support center $\bm{\mu}_{\mathrm{tr}}$ and regularized precision matrix $\bm{\Lambda}_{\mathrm{tr}}$, the comparison-level distance is
\begin{equation}
D_t
=
\max_{j}
\sqrt{
(\bm{z}_{t,j}-\bm{\mu}_{\mathrm{tr}})^{\mathsf{T}}
\bm{\Lambda}_{\mathrm{tr}}
(\bm{z}_{t,j}-\bm{\mu}_{\mathrm{tr}})
}.
\label{eq:joint-support-distance}
\end{equation}
Taking the maximum requires every candidate in the released comparison to remain supported. The horizon $H$ is inherited directly from the query and is checked against the longest horizon whose rollout error was validated.

Let $q_U(H)$ denote the horizon-specific uncertainty threshold calibrated on validation trajectories, $q_D$ the radius of the training-support region, and $H^{\mathrm{cal}}$ the longest validated horizon. The decision is
\begin{equation}
d_t
=
\begin{cases}
\mathrm{accept},
& U_t\leq q_U(H),\ D_t\leq q_D,\ H\leq H^{\mathrm{cal}},\\
\mathrm{abstain},
& \text{otherwise}.
\end{cases}
\label{eq:trust-gate-decision}
\end{equation}
Under this decision, an accepted query releases $\mathcal{Z}_t$ through \eqref{eq:starworld-output}; an abstained query returns whether uncertainty, joint support, or horizon caused rejection. The decision therefore applies directly to the same futures and configuration effects produced by Cross-Layer Imagination.

\subsection{Learning and Calibration}

Learning and calibration are the two offline stages that construct StarWorld's query-time components. Model learning equips Cross-Layer Imagination with probabilistic network dynamics and a playback correction, while preserving the predictor, controller, and analytical playback process as executable operations. Trust calibration then equips the Trust Gate with a training-support region, horizon-specific uncertainty thresholds, and a validated rollout horizon. Once these stages finish, their parameters remain frozen during a query.

\subsubsection{Model Learning}

Historical Starlink measurements, aligned prediction records, and streaming replays are organized into causal cross-layer trajectories. From these trajectories, StarWorld learns the network dynamics $\mathcal{F}_{\theta}$ used in \eqref{eq:probabilistic-network-futures} and the playback correction $\delta_{\phi}$ used in \eqref{eq:executable-cross-layer-rollout}. The configured predictor, controller, and analytical playback transition are executed rather than replaced by additional learned surrogates.

The training objective is summarized as
\begin{equation}
\mathcal{L}
=
\mathcal{L}_{\mathrm{future}}
+
\lambda_C\mathcal{L}_{\mathrm{cross}}
+
\lambda_H\mathcal{L}_{\mathrm{multi}},
\label{eq:starworld-training-objective}
\end{equation}
The three terms in \eqref{eq:starworld-training-objective} fit the probabilistic network distribution with emphasis on lower-tail behavior, supervise prediction and playback consequences, and constrain recursive drift over multiple steps, respectively.

For the playback correction, multi-step supervision compares rolled-out buffer and cumulative rebuffering with replay states. This also allows the correction to be refined with the network and prediction components held fixed, preserving the upstream dynamics and executable interfaces.

\subsubsection{Trust Calibration}

Trust calibration begins after model fitting. Training queries define $\bm{\mu}_{\mathrm{tr}}$, $\bm{\Lambda}_{\mathrm{tr}}$, and the support radius $q_D$ in \eqref{eq:joint-support-distance}. StarWorld then performs factual rollouts on validation trajectories. For horizon $h$, the validation pairs of uncertainty and realized rollout error are summarized as
\begin{equation}
\left(q_U(h),\varepsilon_h\right)
=
\operatorname{Calibrate}
\left(
\left\{\left(U_i(h),E_i(h)\right)\right\}_{i\in\mathcal{D}_{\mathrm{val}}}
\right),
\label{eq:trust-calibration}
\end{equation}
where $E_i(h)$ is the factual rollout error and $q_U(h)$ is selected so that accepted validation queries satisfy the target error bound $\varepsilon_h$ at the prescribed coverage. The trusted horizon $H^{\mathrm{cal}}$ is the longest calibrated horizon that meets this criterion. Test outcomes participate only in final evaluation.

\subsection{Algorithmic Summary}

The complete StarWorld query combines future generation, executable application rollout, paired comparison, and trust assessment. Algorithm~\ref{alg:starworld-query} summarizes this procedure.

\begin{algorithm}[H]
\caption{StarWorld Cross-Layer What-If Query}
\label{alg:starworld-query}
\begin{algorithmic}[1]
\STATE \textbf{Input:} observed history $\bm{s}_{\leq t}$, context $\bm{c}$, configurations $\mathcal{U}$, horizon $H$, sample count $M$
\FOR{$m=1$ to $M$}
    \STATE Sample a network future using \eqref{eq:probabilistic-network-futures}
    \FOR{each configuration $\bm{u}_{j}\in\mathcal{U}$}
        \STATE Execute the transitions in \eqref{eq:executable-cross-layer-rollout}
        \STATE Store trajectory $\widehat{\tau}_{j}^{(m)}$ from \eqref{eq:cross-layer-imagination}
    \ENDFOR
\ENDFOR
\STATE Aggregate outcome distributions and paired effects
\STATE Construct result package $\mathcal{Z}_t$ in \eqref{eq:imagination-result-package}
\STATE Compute $U_t$ and $D_t$ using \eqref{eq:rollout-uncertainty} and \eqref{eq:joint-support-distance}
\STATE Apply the Trust Gate in \eqref{eq:trust-gate-decision}
\IF{$d_t=\mathrm{accept}$}
    \STATE Release trajectories, outcome distributions, and paired effects
\ELSE
    \STATE Abstain and report the rejection reason
\ENDIF
\STATE \textbf{Output:} trusted cross-layer comparison or abstention
\end{algorithmic}
\end{algorithm}

\section{Experiments}
\label{sec:experiments}

\subsection{Experimental Setting}
\label{sec:experimental-setting}

\subsubsection{Starlink Data and Cross-Layer Episode Archive}

We use real user-side measurements from the Starlink dataset~\cite{liu2025starnet}, collected in Osnabr\"uck, Germany (OSN), Chicago, United States (US), and Victoria, Canada (VIC). Table~\ref{tab:starworld-episode-archive} summarizes their regional coverage, the prediction and replay records derived from them, and the state information available at each layer.

A valid segment defines an episode boundary rather than an individual world-model state. Each segment supplies a measurement--prediction episode; replay-covered segments additionally supply a branch for each controller. Network and prediction states share a common clock, with streaming states aligned using the most recently completed chunk at each boundary. All measurement episodes supervise network dynamics, while replay branches provide application-state and action supervision.

\begin{table}[!t]
\centering
\caption{Starlink measurement data and cross-layer evidence archive.}
\label{tab:starworld-episode-archive}
\small
\setlength{\tabcolsep}{5pt}
\renewcommand{\arraystretch}{1.14}
\renewcommand{\tabularxcolumn}[1]{m{#1}}
\begin{tabularx}{\columnwidth}{@{}>{\raggedright\arraybackslash}m{0.41\columnwidth}>{\raggedright\arraybackslash}X@{}}
\toprule
\textbf{Data component} & \textbf{Extent or content} \\
\midrule
\multicolumn{2}{@{}l}{\textbf{Measurement coverage}} \\
\addlinespace[2pt]
Chicago (US) & 4,421 valid segments \\
Osnabr\"uck (OSN) & 1,150 valid segments \\
Victoria (VIC) & 243 valid segments \\
All regions & 5,814 valid 10-min segments \\
Prediction windows & 313,956 overlapping windows \\
\midrule
\multicolumn{2}{@{}l}{\textbf{Cross-layer replay coverage}} \\
\addlinespace[2pt]
Replay-covered segments & 1,024 \\
Controller replay branches & 4,096 (1,024 per controller) \\
Aligned 10-s replay intervals & 27,745 across controller branches \\
\midrule
\multicolumn{2}{@{}l}{\textbf{Recorded state information}} \\
\addlinespace[2pt]
Measurement layer & Throughput statistics, lower-tail capacity, and volatility \\
\addlinespace[2pt]
Prediction layer & Point estimate, safe capacity, and prediction spread \\
\addlinespace[2pt]
Streaming layer & Bitrate, buffer, rebuffering, download time, and playback progress \\
\bottomrule
\end{tabularx}
\end{table}

\subsubsection{Replay Environment and Evaluation Protocol}

The replay environment uses BOLA, RobustMPC, Pensieve, and SafeSABR~\cite{spiteri2020bola,yin2015control,mao2017neural,xie2026safesabr} over measured Starlink capacity traces. Queries start from observed causal states, including the available playback prefix, rather than restarting the streaming session. Cross-layer fidelity tests use recorded actions, while controller-effect and ranking studies compare branches under a shared measured network future. Table~\ref{tab:starworld-evaluation-protocol} summarizes the replay parameters and evaluation protocol.

All methods within each comparison use the same data partition and query set, with segments, overlapping windows, and controller branches grouped by collection run. Region, profile, time-period, and controller-family holdouts use separate protocols while preserving this isolation. Training data define the support statistics; selected models and calibrated thresholds remain fixed during test evaluation.

\begin{table}[!t]
\centering
\caption{Replay environment and evaluation protocol.}
\label{tab:starworld-evaluation-protocol}
\small
\setlength{\tabcolsep}{5pt}
\renewcommand{\arraystretch}{1.14}
\renewcommand{\tabularxcolumn}[1]{m{#1}}
\begin{tabularx}{\columnwidth}{@{}>{\raggedright\arraybackslash}m{0.41\columnwidth}>{\raggedright\arraybackslash}X@{}}
\toprule
\textbf{Parameter} & \textbf{Setting} \\
\midrule
\multicolumn{2}{@{}l}{\textbf{Streaming replay environment}} \\
\addlinespace[2pt]
Controllers & BOLA, RobustMPC, Pensieve, SafeSABR \\
Video chunk duration & 4~s \\
Bitrate ladder (Mbps) & 3, 8, 15, 30, 60, 120 \\
World-model state interval & 10~s \\
Query initialization & Observed measurement, prediction, and replay prefix \\
\midrule
\multicolumn{2}{@{}l}{\textbf{Data partitioning and selection}} \\
\addlinespace[2pt]
Partition unit & Measurement collection run \\
Train / validation / test & 60\% / 20\% / 20\% by collection run \\
Model selection and calibration & Validation data only \\
\midrule
\multicolumn{2}{@{}l}{\textbf{Rollout and comparison controls}} \\
\addlinespace[2pt]
Evaluation horizons & 10, 30, 60, 120~s \\
Nominal prediction interval & 90\% \\
Low-capacity threshold & 10~Mbps for P05 throughput \\
Controller-effect comparison & Shared factual network continuation across branches \\
Conditional rollout actions & Recorded action at each step \\
\bottomrule
\end{tabularx}
\end{table}

\subsubsection{Comparison Methods}
\label{sec:comparison-methods}

The main network-future comparison includes four statistical methods. \emph{Persistence} repeats the latest network state, whereas \emph{Recent mean} holds the mean of recent observed states fixed. \emph{Autoregressive} fits a ridge-regression transition from recent history to the next state and applies it recursively. \emph{Causal empirical} constructs a regional distribution of observed one-step changes from training data and samples successive increments to generate future paths.

Three neural baselines model temporal dependence and predictive uncertainty. \emph{Probabilistic gated recurrent unit (GRU)}~\cite{cho2014gru} encodes the observed history recurrently and predicts Gaussian state increments, feeding each generated state into the next step. \emph{Causal Transformer}~\cite{vaswani2017attention} uses causal self-attention and Gaussian increment heads to generate futures autoregressively. \emph{Generic recurrent state-space model (RSSM)}~\cite{hafner2019planet} combines a recurrent state with Gaussian latent priors and training-time observation posteriors, learned through multi-step likelihood and Kullback--Leibler regularization; inference uses only the prior.

For cross-layer rollout, \emph{Persistence + equations} holds network capacity fixed and advances playback with the analytical transition. \emph{Per-layer recursive} fits separate one-step ridge transitions in network--prediction--streaming order, while \emph{Joint recursive} fits one transition over the concatenated cross-layer state. Both learned baselines recursively feed back their predicted states, starting from the same observed states and supplied actions as StarWorld.

For controller effects, \emph{Controller lookup} and \emph{Profile lookup} return training-set mean outcomes by controller and by profile--controller pair, respectively. \emph{No action} uses ridge regression to predict terminal outcomes from the initial state and context. \emph{Direct action} additionally receives the controller configuration and supplied action sequence, predicts terminal rebuffering, and computes QoE using that estimate and the same bitrate-based reward definition as StarWorld.

\subsubsection{Implementation Details}
\label{sec:starworld-implementation}

The network-future generator uses a two-layer causal Transformer with hidden width 96, four attention heads, feedforward width 288, and dropout 0.1. Gaussian heads predict state increments and their standard deviations. A bounded state-conditioned correction, fitted with gradient-boosted trees on training data, adjusts the forecast center; a 0.1-weight blend of centered trajectories retrieved from region- and state-matched training histories supplies empirical trajectory support. The learned replay-transition model uses a 96-unit GRU state and a 32-dimensional Gaussian latent, with separate 96-wide multilayer perceptrons for prediction and playback-residual outputs. These branches use SiLU activations, layer normalization, and dropout 0.1; the playback residual corrects the analytical transition.

The neural network-future baselines also use hidden width 96. The GRU has a single recurrent layer, the Causal Transformer uses the same attention configuration as StarWorld's network-future generator, and the RSSM uses 32 latent dimensions. Recent mean and Autoregressive use the latest six states, and all ridge-regression baselines use regularization strength 1.

Features are standardized using training statistics, and checkpoints are selected on validation data. Neural network-future models are trained for up to 20 epochs with AdamW, batch size 512, learning rate $8\times10^{-4}$, weight decay $10^{-5}$, cosine learning-rate decay, gradient-norm clipping at 5, and early-stopping patience of six epochs. Replay-transition training uses the same AdamW learning rate and weight decay, with 15 epochs, batch size 256, and 24-step sequences; the multi-step supervision horizon increases to six steps during training.

In the main network-future comparison, 120~s of observed history conditions 24 sampled trajectories; learned methods use three independently trained members with eight samples each. Deterministic statistical forecasts are converted to distributions using temporally correlated Gaussian perturbations scaled by training-set state increments. Every method then uses validation-only spread calibration and the same nonnegative capacity-ordering constraints.

\subsubsection{Evaluation Metrics}

We evaluate probabilistic network futures using the continuous ranked probability score (CRPS) and its threshold-weighted form (twCRPS), written jointly as
\begin{equation}
\operatorname{CRPS}_{w}(F,y)
=
\int_{-\infty}^{\infty}
w(z)\left(F(z)-\mathbf{1}\{z\geq y\}\right)^2\,\mathrm{d}z,
\label{eq:experimental-crps}
\end{equation}
where $F$ is the predictive cumulative distribution, $y$ is the observation, and $w$ is a weighting function. Setting $w(z)=1$ gives CRPS, applied separately to window-mean throughput and within-window fifth-percentile (P05) throughput. For P05, $w(z)=\mathbf{1}\{z\leq c\}$ with $c=10$~Mbps gives low-capacity twCRPS, emphasizing distributional accuracy below the threshold. Lower scores indicate better fidelity. The 90\% interval coverage is the fraction of observations within the central 90\% predictive interval; values close to 90\% indicate calibration.

Pointwise cross-layer fidelity for state variable $q$ at horizon $H$ is measured by the mean absolute error (MAE),
\begin{equation}
\operatorname{MAE}_{q}(H)
=
\frac{1}{N_H}
\sum_{i=1}^{N_H}
\left|\widehat{q}_{i,t+H}-q_{i,t+H}\right|,
\label{eq:experimental-mae}
\end{equation}
where $q$ denotes throughput, prediction, buffer, or cumulative rebuffering, $\widehat{q}$ is its point estimate, and $N_H$ is the number of eligible queries at horizon $H$. Pairwise-effect MAE applies the same error measure to predicted and observed QoE or rebuffer differences between controller branches.

Playback QoE is $Q=\bar b-0.1\bar d-4.3R/L$, where $\bar b$ is mean interval bitrate weighted by $\max(n_\ell,1)$ for interval decision count $n_\ell$, $\bar d$ is mean absolute bitrate change between intervals, $R$ is total rebuffering in seconds, and $L$ is the number of 10-s intervals.

Trust evaluation uses Pearson correlation between uncertainty and realized rollout error, and risk--coverage curves in which coverage is the retained-query fraction and risk is MAE over those queries. The area under the receiver operating characteristic curve (AUROC) measures discrimination between supported and unsupported queries, with unsupported queries as the positive class.

\subsection{Probabilistic Network-Future Fidelity}
\label{sec:multistep-world-fidelity}

This experiment compares StarWorld with statistical and learned sequence models to evaluate the fidelity of its generated network futures. We examine both overall capacity distributions and lower-tail behavior, which provide the network-dynamics foundation for subsequent cross-layer rollouts.

We compare StarWorld with seven network-future baselines at 30 and 60~s using throughput CRPS, P05 CRPS, low-capacity twCRPS, and 90\% interval coverage. Table~\ref{tab:starworld-endpoint-fidelity} uses T for throughput, LC-twCRPS for low-capacity twCRPS, and Hor. and Cov. for horizon and coverage.

\begin{figure*}[t]
  \centering
  \includegraphics[width=0.93\textwidth]{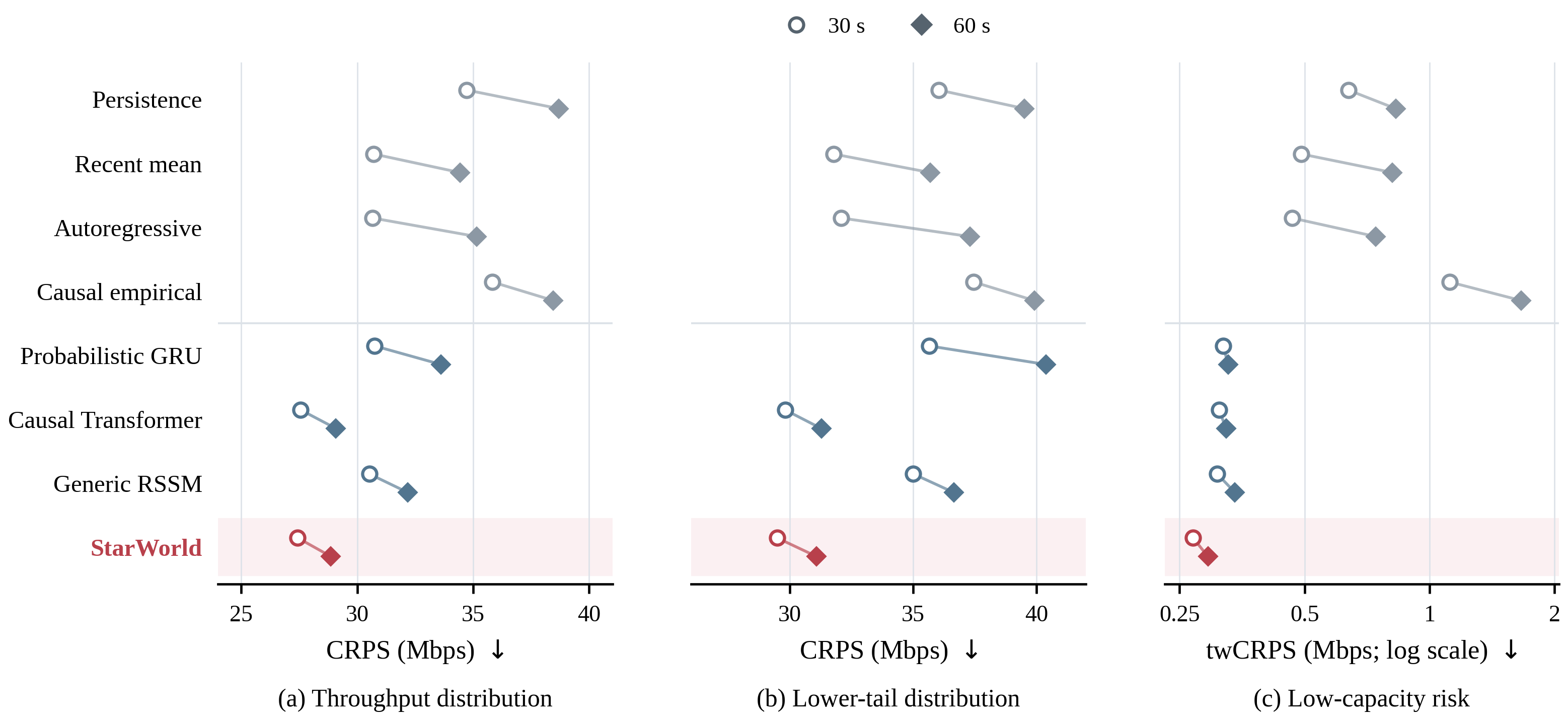}
  \caption{Test-set network-future fidelity at 30 and 60~s: (a) throughput CRPS; (b) P05 CRPS; (c) low-capacity twCRPS (log scale).}
  \label{fig:starworld-multistep-fidelity}
\end{figure*}

\begin{table}[!t]
\centering
\caption{Test-set endpoint fidelity and interval coverage.}
\label{tab:starworld-endpoint-fidelity}
\small
\setlength{\tabcolsep}{3pt}
\renewcommand{\arraystretch}{1.15}
\begin{tabular*}{\columnwidth}{@{\extracolsep{\fill}}clcccc@{}}
\toprule
\multirow{2}{*}{\textbf{Hor.}} &
\multirow{2}{*}{\textbf{Method}} &
\multicolumn{2}{c}{\textbf{CRPS} $\downarrow$} &
\multirow{2}{*}{\shortstack{\textbf{LC-}\\\textbf{twCRPS} $\downarrow$}} &
\multirow{2}{*}{\shortstack{\textbf{90\%}\\\textbf{Cov.}}} \\
\cmidrule(lr){3-4}
 & & \textbf{T} & \textbf{P05} & & \\
\midrule
\multirow{8}{*}{30~s}
 & Persistence & 34.725 & 36.045 & 0.637 & 85.9\% \\
 & Recent mean & 30.715 & 31.777 & 0.491 & 66.8\% \\
 & Autoregressive & 30.662 & 32.079 & 0.466 & 65.6\% \\
 & Causal empirical & 35.825 & 37.455 & 1.117 & 58.7\% \\
 & Probabilistic GRU & 30.756 & 35.661 & 0.319 & 69.4\% \\
 & Causal Transformer & 27.567 & 29.808 & 0.311 & \textbf{90.8\%} \\
 & Generic RSSM & 30.521 & 35.007 & 0.308 & 71.6\% \\
 & StarWorld & \textbf{27.436} & \textbf{29.482} & \textbf{0.269} & 79.9\% \\
\midrule
\multirow{8}{*}{60~s}
 & Persistence & 38.690 & 39.517 & 0.829 & 65.1\% \\
 & Recent mean & 34.438 & 35.703 & 0.814 & 69.8\% \\
 & Autoregressive & 35.146 & 37.303 & 0.742 & 65.0\% \\
 & Causal empirical & 38.457 & 39.915 & 1.662 & 65.5\% \\
 & Probabilistic GRU & 33.619 & 40.398 & 0.327 & 74.5\% \\
 & Causal Transformer & 29.086 & 31.275 & 0.324 & 91.6\% \\
 & Generic RSSM & 32.180 & 36.647 & 0.340 & 92.1\% \\
 & StarWorld & \textbf{28.865} & \textbf{31.077} & \textbf{0.292} & \textbf{89.1\%} \\
\bottomrule
\end{tabular*}
\end{table}

Fig.~\ref{fig:starworld-multistep-fidelity}(a) and (b) compare overall throughput and lower-tail distributions across all methods. StarWorld obtains the lowest CRPS for both quantities at both horizons. Table~\ref{tab:starworld-endpoint-fidelity} gives the exact values: at 60~s, its throughput CRPS is 28.865, compared with 29.086 for Causal Transformer, 32.180 for Generic RSSM, and 33.619 for Probabilistic GRU; the statistical baselines range from 34.438 to 38.690. Its P05 CRPS of 31.077 is also the lowest. These results show improved fidelity for both typical and lower-tail capacity, with Causal Transformer providing the closest comparison.

The low-capacity comparison in Fig.~\ref{fig:starworld-multistep-fidelity}(c) shows a clearer separation from the statistical baselines. At 60~s, their twCRPS values range from 0.742 to 1.662, while the three neural baselines range from 0.324 to 0.340. StarWorld reaches 0.292 and also ranks first at 30~s with 0.269. Its advantage therefore extends to the rare low-capacity futures that are particularly relevant to playback risk.

Interval coverage complements this accuracy comparison. At 60~s, StarWorld reaches 89.1\% coverage for the nominal 90\% interval, whereas its 30-s coverage is 79.9\%; Causal Transformer is better calibrated at the shorter horizon. Together, the all-method comparison establishes a more faithful network-future foundation for Cross-Layer Imagination, with the strongest evidence in lower-tail and rare-capacity modeling.

\subsection{Executable Cross-Layer Rollout Fidelity}
\label{sec:cross-layer-rollout-fidelity}

This experiment evaluates whether Cross-Layer Imagination preserves buffer and cumulative-rebuffer states when sampled network futures are propagated through the executable application pipeline.

We compare StarWorld with Persistence + equations, Per-layer recursive, and Joint recursive at 30, 60, and 120~s using buffer and cumulative-rebuffer MAE. Only the playback correction is refined for 50 epochs with AdamW (learning rate $3\times10^{-4}$, batch size 128), keeping upstream parameters fixed and selecting checkpoints on validation data. State estimates average eight futures from each of three stochastic members.

\begin{figure*}[t]
  \centering
  \includegraphics[width=0.85\textwidth]{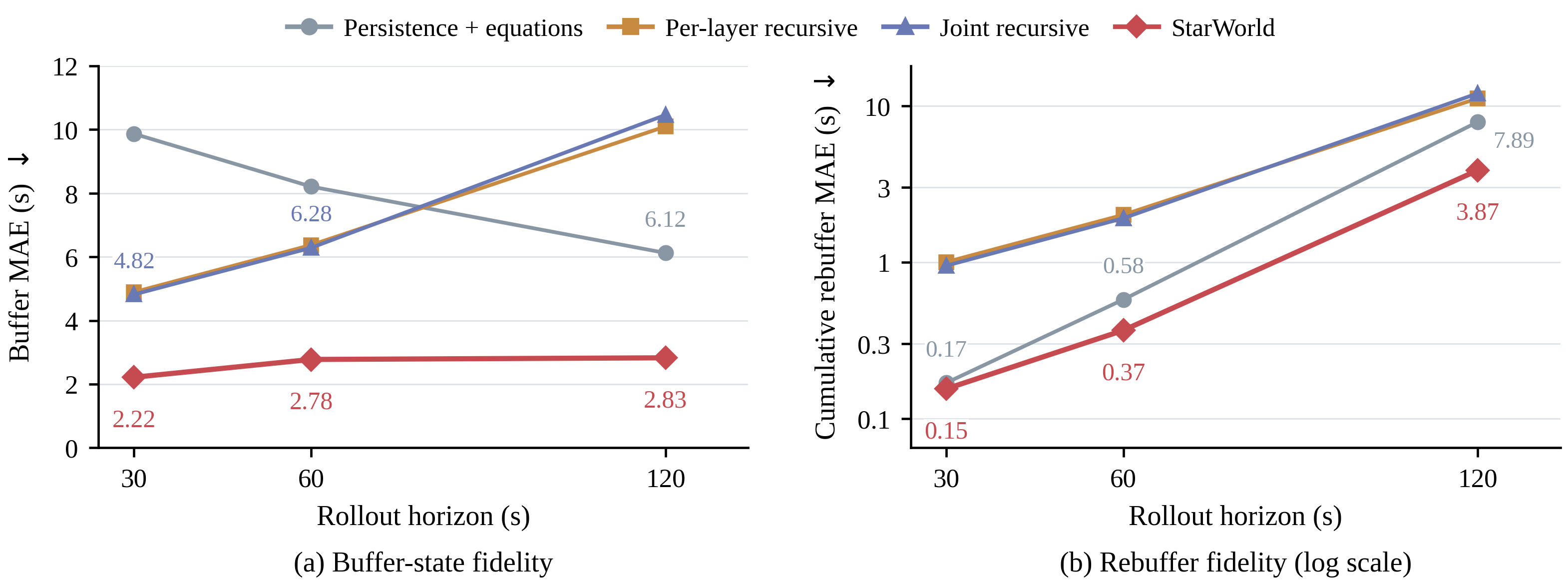}
  \caption{Cross-layer rollout fidelity: (a) buffer MAE; (b) cumulative-rebuffer MAE (log scale).}
  \label{fig:starworld-cross-layer-fidelity}
\end{figure*}

Fig.~\ref{fig:starworld-cross-layer-fidelity}(a) shows that StarWorld achieves the lowest buffer MAE at every evaluated horizon. Its errors are 2.22, 2.78, and 2.83~s at 30, 60, and 120~s, respectively, compared with 4.82, 6.28, and 6.12~s for the strongest baseline at each horizon. Joint recursive is the closest competitor at 30 and 60~s, while Persistence + equations is strongest at 120~s. The buffer-state advantage therefore holds against both learned recursion and analytical persistence.

Fig.~\ref{fig:starworld-cross-layer-fidelity}(b) tests whether this advantage also extends to cumulative rebuffering. Persistence + equations is the strongest baseline at all three horizons, with errors of 0.17, 0.58, and 7.89~s. StarWorld lowers these errors to 0.15, 0.37, and 3.87~s, respectively. The difference is modest at 30~s but widens at 60 and 120~s, showing that the buffer improvement does not come at the cost of less accurate rebuffer estimates.

Together, these results show that StarWorld propagates network futures into more faithful playback states than either analytical persistence or learned recursive transitions on the evaluated queries.

\subsection{Shared-Future Controller-Effect Accuracy}
\label{sec:action-counterfactual-accuracy}

This experiment evaluates how accurately StarWorld estimates outcome differences among controller branches under a shared network future. Holding the factual network continuation fixed across all branches isolates controller-effect accuracy from network-future error.

We evaluate 138 test trace groups with complete four-controller coverage over shared 60-s measured futures, focusing on 89 with nontrivial QoE or rebuffer differences. Metrics are pairwise-effect MAE, sign accuracy for absolute observed differences above 0.05, Spearman controller-rank correlation, and top-1 agreement with the replay-best QoE controller.

\begin{figure*}[t]
  \centering
  \includegraphics[width=0.93\textwidth]{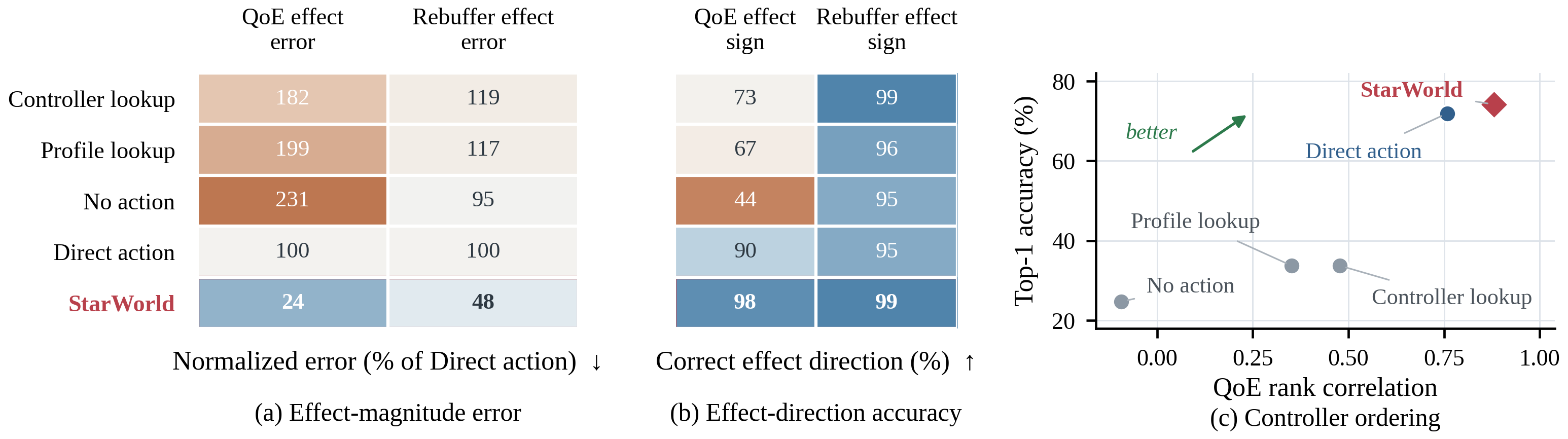}
  \caption{Shared-future controller effects on action-sensitive traces: (a) pairwise errors normalized by Direct action; (b) sign accuracy; (c) QoE rank correlation and top-1 accuracy.}
  \label{fig:starworld-action-counterfactual}
\end{figure*}

Fig.~\ref{fig:starworld-action-counterfactual}(a) shows the clearest difference in effect magnitude. With Direct action normalized to 100\%, StarWorld reduces QoE-effect error to 24\% and rebuffer-effect error to 48\%. Executing each controller branch step by step therefore captures its outcome difference more accurately than a one-shot terminal mapping.

The predicted effects also retain their direction. In Fig.~\ref{fig:starworld-action-counterfactual}(b), StarWorld reaches 98\% QoE-sign accuracy and 99\% rebuffer-sign accuracy. Removing action information lowers QoE-sign accuracy to 44\%, showing that the initial network profile alone does not explain the observed controller differences.

Fig.~\ref{fig:starworld-action-counterfactual}(c) connects these pairwise effects to controller ordering. StarWorld occupies the strongest upper-right operating point, exceeding Direct action in both QoE rank correlation and top-1 accuracy, whereas the no-action model has a negative rank correlation. Together, the magnitude, direction, and ordering results validate paired controller effects under shared network conditions.

\subsection{Generalization Boundary of Cross-Layer Imagination}
\label{sec:cross-domain-boundary}

This experiment characterizes where Cross-Layer Imagination retains useful streaming-state fidelity under unseen regions, profiles, time periods, controller actions, and long-horizon conditions. These holdouts expose the domain and horizon boundaries that the Trust Gate must distinguish from supported queries.

We construct eight evaluation settings: an in-distribution reference; US, OSN, and VIC region holdouts; a low-tail-profile holdout; a chronological future-period holdout; a SafeSABR action-family holdout; and a long-horizon stress set. StarWorld is compared with the per-layer recursive baseline using cumulative-rebuffer MAE at 30, 60, and 120~s, with relative improvement defined as the percentage reduction from baseline MAE.

\begin{figure*}[t]
  \centering
  \includegraphics[width=0.93\textwidth]{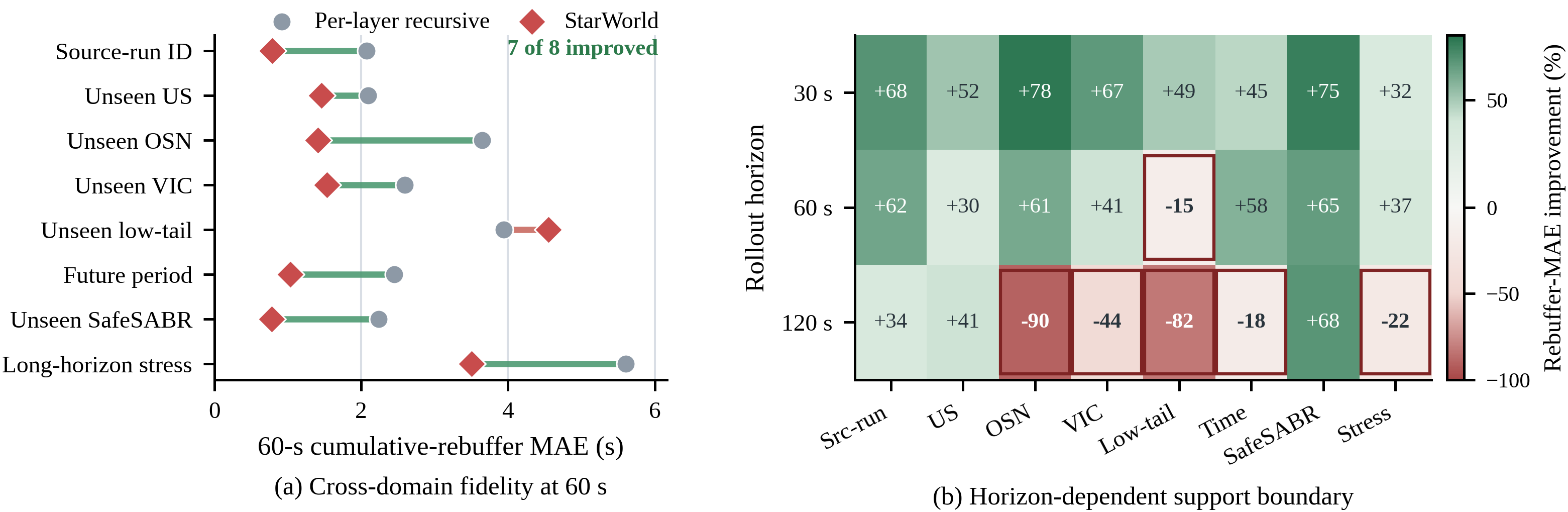}
  \caption{Generalization boundary: (a) 60-s cumulative-rebuffer MAE; (b) relative MAE reduction over Per-layer recursive across horizons.}
  \label{fig:starworld-cross-domain-boundary}
\end{figure*}

Fig.~\ref{fig:starworld-cross-domain-boundary}(a) shows that StarWorld lowers 60-s cumulative-rebuffer MAE in seven of the eight settings. The corresponding 60-s row in Fig.~\ref{fig:starworld-cross-domain-boundary}(b) reports improvements of 30\% for US, 61\% for OSN, 41\% for VIC, and 58\% for the future-period holdout. The low-tail holdout is the only reversal, with a $-15\%$ improvement.

The horizon view in Fig.~\ref{fig:starworld-cross-domain-boundary}(b) separates transfer from extrapolation. StarWorld improves all eight settings at 30~s and seven at 60~s, but only three remain favorable at 120~s. The SafeSABR holdout is notably stable despite removing that controller family from training, with gains of 65\% and 68\% at 60 and 120~s. In contrast, the low-tail, OSN, VIC, chronological, and long-horizon stress settings reverse at 120~s.

These results establish a bounded generalization claim: StarWorld transfers its streaming-state advantage across diverse domains at short and medium horizons, including an unseen action family, while lower-tail and long-horizon shifts expose explicit support limits.

\subsection{Trust Gate Reliability}
\label{sec:world-model-trustworthiness}

This experiment evaluates whether the Trust Gate recognizes unreliable imagined comparisons. It tests both whether stochastic uncertainty ranks factual rollout error and whether joint support detects unsupported cross-layer queries that preserve a familiar network trajectory.

We evaluate the three-member ensemble at 30, 60, and 120~s and compare accept-all, uncertainty-only, network-support-only, and the full Trust Gate. A controlled support stress keeps the measured network state fixed while introducing unsupported controller and context combinations, separating joint-query support from conventional network-only detection.

\begin{figure*}[t]
  \centering
  \includegraphics[width=0.93\textwidth]{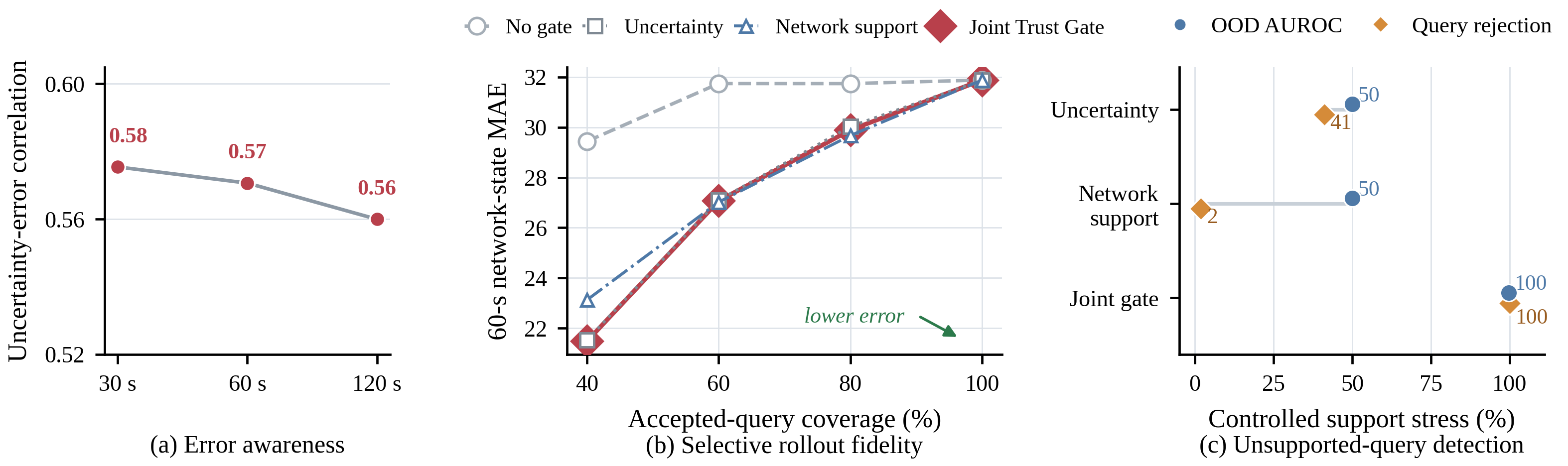}
  \caption{Trust Gate reliability: (a) uncertainty--error correlation; (b) 60-s risk--coverage tradeoff; (c) unsupported-query detection with unchanged network states.}
  \label{fig:starworld-trustworthiness}
\end{figure*}

Fig.~\ref{fig:starworld-trustworthiness}(a) shows that uncertainty remains positively associated with factual rollout error across horizons: the displayed correlations are 0.58, 0.57, and 0.56 at 30, 60, and 120~s. The ensemble spread therefore provides a stable reliability ranking even as the rollout horizon grows.

This ranking translates into selective fidelity. In Fig.~\ref{fig:starworld-trustworthiness}(b), the 60-s full Trust Gate lowers network-state MAE as coverage decreases and remains below unguided selection at matched coverage. Uncertainty-only selection follows a similar in-distribution curve, indicating that stochastic uncertainty drives most of the error ranking on supported test queries.

The controlled stress in Fig.~\ref{fig:starworld-trustworthiness}(c) reveals the complementary role of joint support. Because the network trajectory is unchanged, uncertainty-only and network-support checks remain at 50\% AUROC, whereas the joint Trust Gate reaches nearly 100\% AUROC and rejects all unsupported queries. Together, these results validate the uncertainty and joint-support signals used by the Trust Gate to distinguish high-error rollouts from unsupported cross-layer comparisons.

\subsection{Trace-Conditioned Configuration Ranking}
\label{sec:trace-conditioned-ranking}

\begin{figure*}[t]
  \centering
  \includegraphics[width=0.93\textwidth]{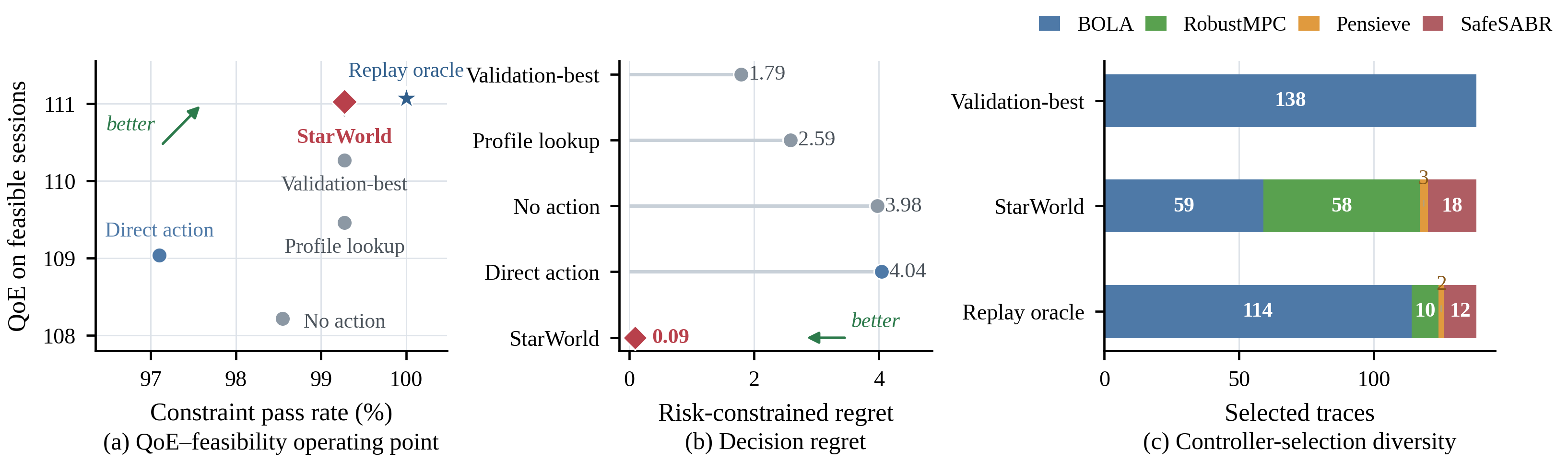}
  \caption{Offline configuration ranking: (a) feasible-session QoE versus constraint pass rate; (b) risk-constrained regret; (c) controller-selection composition.}
  \label{fig:starworld-controller-ranking}
\end{figure*}

This experiment examines whether the paired outcome estimates validated above support offline controller ranking under a playback-risk constraint. Holding the measured network future common across all branches makes ranking quality reflect the configuration-effect estimates rather than differences in network conditions.

On 138 test traces with shared 60-s measured futures, the selector chooses the highest-QoE predicted-feasible controller under a 0.5-s rebuffer limit, or the minimum-risk controller otherwise. Comparisons include a validation-best fixed controller and a constraint-first replay oracle. Metrics are constraint pass rate, mean QoE among feasible selections, and mean risk-constrained regret $[Q^{\star}-Q]_+ +10[R-0.5]_+$, where $Q^{\star}$ is oracle QoE and $[x]_+=\max(x,0)$.

Fig.~\ref{fig:starworld-controller-ranking}(a) shows that StarWorld retains the high constraint pass rate of the validation-best controller while moving feasible-session QoE close to the replay-oracle operating point. This places StarWorld near the upper-right boundary of the QoE--feasibility comparison.

The larger difference appears in decision regret. In Fig.~\ref{fig:starworld-controller-ranking}(b), StarWorld reduces risk-constrained regret to 0.09, compared with 1.79 for validation-best and 4.04 for the direct-action model. The substantially lower regret shows that the outcome estimates preserve the relative value of feasible alternatives rather than only their average quality.

Fig.~\ref{fig:starworld-controller-ranking}(c) confirms that this gain is not produced by selecting one hidden fixed policy. Validation-best uses BOLA on all traces, whereas StarWorld distributes its selections across all four controller families and approaches the replay oracle's operating point. StarWorld therefore turns shared-future effect fidelity into useful offline ranking over the covered controller configurations.

\subsection{Efficiency and Scalability}
\label{sec:world-model-efficiency}

\begin{figure*}[t]
  \centering
  \includegraphics[width=0.93\textwidth]{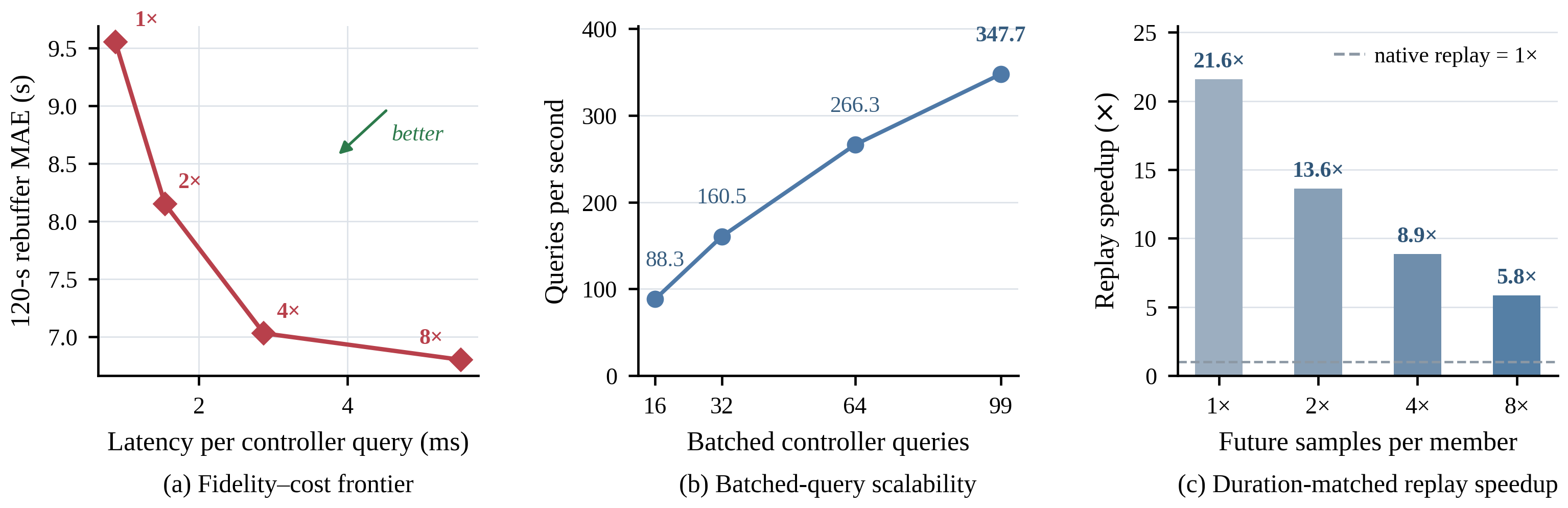}
  \caption{Efficiency and scalability: (a) 120-s fidelity--cost tradeoff; (b) batched-query throughput; (c) amortized trajectory speedup over duration-matched BOLA replay.}
  \label{fig:starworld-efficiency}
\end{figure*}

\begin{figure*}[t]
  \centering
  \includegraphics[width=0.93\textwidth]{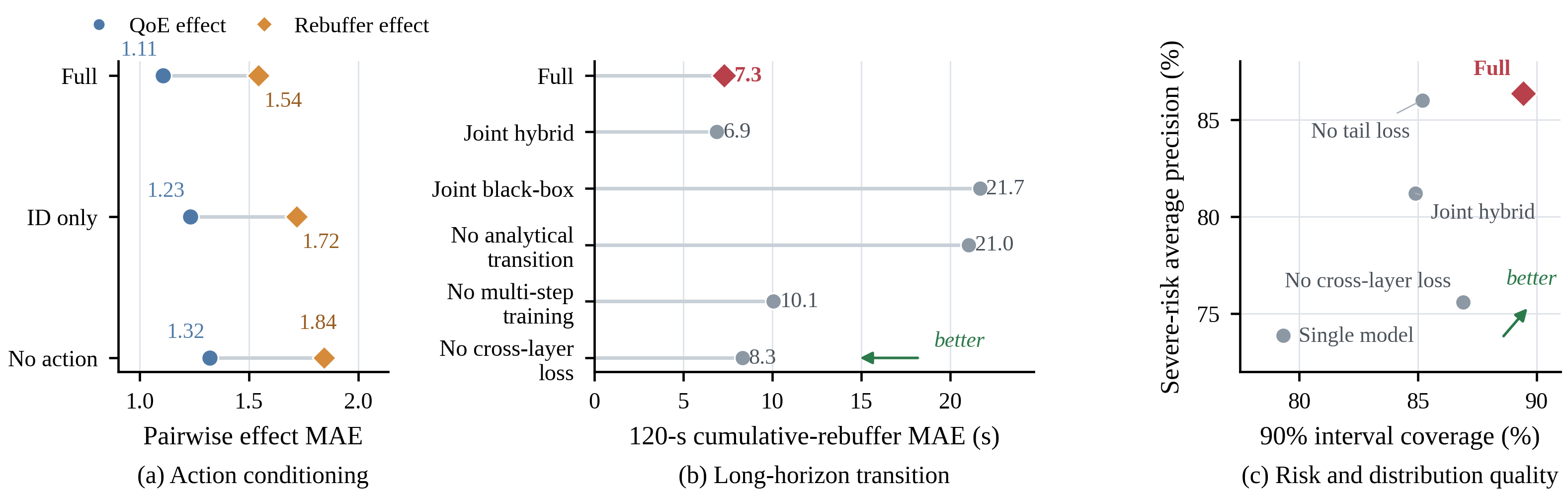}
  \caption{Module ablation: (a) controller-effect MAE; (b) 120-s cumulative-rebuffer MAE; (c) severe-risk average precision and 90\% interval coverage.}
  \label{fig:starworld-module-ablation}
\end{figure*}

This experiment evaluates whether Cross-Layer Imagination can amortize future evaluation across stochastic samples and batched configuration queries. It measures the fidelity--cost tradeoff of imagined rollouts and their practical wall-clock cost relative to duration-matched native replay.

With two ensemble members, we evaluate 30-, 60-, and 120-s rollouts using 1, 2, 4, or 8 future samples per member; query scaling is measured at 120~s with four samples. The duration-matched native BOLA reference executes 30 four-second chunks. We report latency per configuration query, throughput in queries/s, and amortized trajectory speedup, defined as native replay time per trajectory divided by StarWorld time per generated future.

Fig.~\ref{fig:starworld-efficiency}(a) exposes a controllable fidelity--cost tradeoff: additional future samples progressively lower 120-s rebuffer MAE at the cost of higher latency per configuration query. Four samples lie near the knee of this frontier and provide an intermediate operating point.

Batching further amortizes the query workload. Fig.~\ref{fig:starworld-efficiency}(b) shows that throughput rises from 88.3 queries/s for a 16-query batch to 347.7 queries/s for 99 queries, rather than scaling as repeated serial calls. Fig.~\ref{fig:starworld-efficiency}(c) gives the corresponding duration-matched comparison: depending on the requested number of futures, each imagined trajectory is 5.8--21.6 times faster than native replay.

StarWorld therefore supports many stochastic cross-layer queries at millisecond-scale latency while allowing the user to trade computation for rollout fidelity. This makes Cross-Layer Imagination an amortized evaluator for configuration exploration and ranking, with native replay retained as the final accuracy reference.

\subsection{Module Ablation}
\label{sec:starworld-ablation}

This experiment identifies which lower-level mechanisms support Cross-Layer Imagination and the reliability signals used by the Trust Gate. Each mechanism is evaluated using the capability it directly affects: paired controller effects, long-horizon streaming transitions, or probabilistic risk quality.

All trainable variants, including a retrained Full model, share the same 15-epoch budget and model scale. Ablations cover action representation, transition structure, training objectives, and ensemble uncertainty. At 60~s, probabilistic variants report 90\% interval coverage and average precision across recall levels for severe-risk detection (cumulative rebuffering $\geq10$~s).

Fig.~\ref{fig:starworld-module-ablation}(a) shows that learned action conditioning improves effect magnitude beyond controller identity. Removing action features raises QoE-effect MAE from 1.11 to 1.32 and rebuffer-effect MAE from 1.54 to 1.84~s, while retaining only controller identity produces intermediate errors of 1.23 and 1.72~s.

The long-horizon result in Fig.~\ref{fig:starworld-module-ablation}(b) isolates the executable transition. Removing the analytical streaming transition raises 120-s cumulative-rebuffer MAE from 7.3 to 21.0~s, while a joint black-box transition reaches 21.7~s. Removing multi-step training and cross-layer supervision increases the error to 10.1 and 8.3~s, respectively. The joint hybrid model reaches 6.9~s, slightly below the factorized full model on this metric; factorization is therefore not credited with improving every streaming error, whereas the analytical transition clearly controls long-horizon rebuffer drift.

Fig.~\ref{fig:starworld-module-ablation}(c) separates risk and distribution quality. The full model occupies the strongest upper-right operating point. A single model lowers both severe-risk average precision and interval coverage; removing the tail loss primarily lowers coverage, while removing cross-layer supervision produces the largest precision decline among the loss variants. These results connect the ensemble to uncertainty quality, the tail objective to calibration, and cross-layer supervision to severe-risk discrimination. Overall, learned action representation supports paired effects, the executable transition limits rollout drift, and probabilistic training supplies the risk and uncertainty signals used by Cross-Layer Imagination and the Trust Gate.

\section{Conclusion}
This paper introduced StarWorld, a cross-layer world model that represents uncertain Starlink access evolution and carries each plausible future through executable prediction and streaming dynamics. Its Cross-Layer Imagination forms connected distributions of measurement, prediction, and playback trajectories, while the Trust Gate identifies the domain and horizon over which those imagined trajectories remain reliable. Experiments on real Starlink measurements show that StarWorld improves lower-tail future fidelity and preserves application-state trajectories and relative controller effects at supported horizons; the cross-domain evaluation also exposes where these advantages reverse. These results demonstrate that a hybrid learned--executable world model can serve as a reliable imagination environment for studying and comparing Starlink prediction and streaming behavior. Future work will incorporate richer orbital, weather, and load context, support online adaptation to evolving access dynamics, and extend the model to settings in which network control and application actions interact bidirectionally.


\end{document}